\documentclass[3p,twocolumn]{elsarticle}

\usepackage{graphicx}
\usepackage{xcolor}
\usepackage{amsmath}
\usepackage{amssymb}

\def \Rey  {\mbox{Re}}
\def \Key  {\mbox{K}}
\def \Ca  {\mbox{Ca}}
\def \muf {\mu^{\rm f}}
\def \mus {\mu^{\rm s}}
\def \sigmaf {\sigma^{\rm f}}
\def \sigmas {\sigma^{\rm s}}
\def \vf {u^{\rm f}}

\def \Phie {\Phi^{\rm e}}
\def \phis {\phi}
\def \NSI {\mathcal{N}_1}
\def \NSII {\mathcal{N}_2}
\def \Tay {\mathcal{T}}

\newcommand{\bra}[1]{\langle #1\rangle}

\newcommand{\ie}{i.e.,\ }
\newcommand{\eg}{e.g.,\ }

\newcommand{\secref}[1]{\mbox{section \ref{#1}}}

\newcommand{\equref}[1]{\mbox{equation (\ref{#1})}}

\newcommand{\tabref}[1]{\mbox{table \ref{#1}}}

\newcommand{\figrefSC}[1]{\mbox{Figure \ref{#1}}}
\newcommand{\figrefS}[1]{\mbox{figure \ref{#1}}}

\begin{document}
\journal{Journal of Non-Newtonian Fluid Mechanics}

\begin{frontmatter}
\title{Suspensions of deformable particles in a Couette flow}
\author{Marco E. Rosti and Luca Brandt}
\address{Linn\'{e} Flow Centre and SeRC (Swedish e-Science Research Centre), \\KTH Mechanics, SE 100 44 Stockholm, Sweden}

\begin{abstract}
We consider suspensions of deformable particles in a Newtonian fluid by means of fully Eulerian numerical simulations with a one-continuum formulation. We study the rheology of the visco-elastic suspension in plane Couette flow in the limit of vanishing inertia and examine the dependency of the effective viscosity $\mu$ on the solid volume-fraction $\Phi$, the capillary number $\Ca$, and the solid to fluid viscosity ratio $\Key$. The suspension viscosity decreases with deformation and applied shear (shear-thinning) while still increasing with volume fraction. We show that $\mu$ collapses to an universal function, $\mu \left( \Phie \right)$, with an effective volume fraction $\Phie$, lower than the nominal one owing to the particle deformation. This universal function is well described by the Eilers fit, which well approximate the rheology of suspension of rigid spheres at all $\Phi$. We provide a closure for the effective volume fraction $\Phie$ as function of volume fraction $\Phi$ and capillary number $\Ca$ and demonstrate it also applies to data in literature for suspensions of capsules and red-blood cells. In addition, we show that the normal stress differences exhibit a non-linear behavior, with a similar trend as in polymer and filament suspensions. The total stress budgets reveals that the particle-induced stress contribution increases with the volume fraction $\Phi$ and decreases with deformability.
\end{abstract}

\end{frontmatter}

\section{Introduction}
Particles suspended in a carrier fluid can be found in many biological, geophysical and industrial flows. Some examples are the blood flow in the human body, pyroclastic flows from volcanoes, sedimentation in sea beds, fluidized beds and slurry flows. Despite the numerous applications, it is still difficult to estimate the force needed to drive suspensions, while in a single phase flow the pressure drop can be accurately predicted as a function of the Reynolds number \cite{pope_2001a} and the properties of the wall surface (e.g.\ roughness \cite{orlandi_leonardi_2008a}, porosity \cite{breugem_boersma_uittenbogaard_2006a, rosti_cortelezzi_quadrio_2015a}, elasticity \cite{rosti_brandt_2017a}). This is due to the complexity of multiphase flows where additional parameters become relevant, such as the size and shape of particles, the density difference with the carrier fluid, their elasticity and the solid volume fraction, denoted here $\Phi$; each of these parameters may be important and affect the overall dynamics of the suspension in different and sometimes surprising ways \cite{mewis_wagner_2012a}. Here, we focus on suspensions where particles are deformable, the solid volume fraction is finite and we use numerical simulations to fully resolve the fluid-structure interactions and the stresses in the solid and liquid. Indeed, there has been growing interest in the study of particles whose shape can modify and adapt to the kind of flow. Examples range from a single liquid droplet with constant surface tension to red blood cells enclosed by a biological membrane or cells with stiff nuclei \cite{freund_2014a, takeishi_imai_ishida_omori_kamm_ishikawa_2016a, alizad-banaei_loiseau_lashgari_brandt_2017a}. Such studies are motivated by the practical need to analyse the behaviour of droplets with contaminated interfaces, cells enclosed by biological membranes, and various synthetic capsules encountered in chemical and biochemical industries, with specific applications in chemical and biomedical engineering.

From a theoretical point of view, Einstein \cite{einstein_1956a} showed in his pioneering work that in the limit of vanishing inertia and for dilute suspensions (i.e.\ $\Phi \rightarrow 0$)
the relative increase in effective viscosity of a suspension of rigid particles in a Newtonian fluid, \ie the suspension viscosity, is a linear function of the particle volume fraction $\Phi$ . Batchelor \cite{batchelor_1977a} and Batchelor and Green \cite{batchelor_green_1972a} added a second order correction in $\Phi$; for higher volume fractions, the viscosity starts to increase faster than a second order polynomial \cite{stickel_powell_2005a}, existing analytical relations are not valid and one needs to resort to empirical fits. One of the available empirical relations for the effective viscosity of rigid particle suspension that provide a good description of the rheology at zero Reynolds number both for the high and low concentration limits is the Eilers fit \cite{ferrini_ercolani_de-cindio_nicodemo_nicolais_ranaudo_1979a, zarraga_hill_leighton-jr_2000a, singh_nott_2003a, kulkarni_morris_2008a}. Inertia has been shown to introduce deviations from the behavior predicted by the different empirical fits, an effect that can be related to an increase of the effective volume fraction at intermediate values of $\Phi$\cite{picano_breugem_mitra_brandt_2013a}. 
Interesting phenomena are also observed at high volume fractions once friction forces become important. 

In this context, understanding the rheology of deformable objects has been a challenge for many years. Deformability is here characterised in terms of the Capillary number $\Ca$, which is the ratio between viscous and elastic forces, so that low $\Ca$'s correspond to configurations dominated by elastic forces where particles easily recover the equilibrium shape and deformations are small. The first efforts to predict the rheological properties of such suspensions is the work by Taylor \cite{taylor_1932a} who assumed small deformations and showed that for small $\Phi$ the coefficient of the linear term in Einstein's relation is a function of the ratio between the particle and fluid viscosities. Later analytical calculations \cite{cox_1969a, frankel_acrivos_1970a, choi_schowalter_1975a, pal_2003a} attempted to extend the result to higher order in $\Phi$ and $\Ca$, similarly to what done by Batchelor for rigid particles, using perturbative expansions. Only recently high-fidelity numerical simulations have been used to study such problem \cite{ii_gong_sugiyama_wu_huang_takagi_2012a, kruger_kaoui_harting_2014a, oliveira_cunha_2015a, srivastava_malipeddi_sarkar_2016a, matsunaga_imai_yamaguchi_ishikawa_2016a}.

In the present work, we will focus on deformable particles with a viscous hyper-elastic behavior. These are a special class of elastic materials (the constitutive behavior is only a function of the current state of deformation) where the work done by the stresses during a deformation process is dependent only on the initial and final configurations, the behaviour of the material is path independent and a stored strain energy function or elastic potential can be defined \cite{bonet_wood_1997a}; these can show nonlinear stress-strain curves and are generally used to describe rubber-like substances. Also, many researchers used materials with similar constitutive relations to simulate particles, capsules, vesicles and even red blood cells  \cite{sugiyama_ii_takeuchi_takagi_matsumoto_2011a, ii_sugiyama_takeuchi_takagi_matsumoto_2011a, ii_gong_sugiyama_wu_huang_takagi_2012a, villone_hulsen_anderson_maffettone_2014a, villone_greco_hulsen_maffettone_2016a}.

\subsection{Outline}
In this work, we present Direct Numerical Simulations (DNS) of a suspension of hyper elastic deformable spheres in a Couette flow at low Reynolds number. The fluid is Newtonian and satisfies the full incompressible Navier-Stokes equations, while momentum conservation and the incompressibility constraint are enforced inside the solid objects. In \secref{sec:formulation}, we first discuss the flow configuration and governing equations, and then present the numerical methodology used. The rheological study of the suspension is presented in \secref{sec:result}, where we also discuss the role of the different parameters defining the elastic particles. We will present a new closure for the shear stress of suspensions of deformable objects, based on an estimate of their deformation, obtained using available numerical and experimental data and the Eilers fit. Finally, a summary of the main findings and some conclusions are drawn  in \secref{sec:conclusion}.

\section{Formulation} \label{sec:formulation}
\begin{figure}
  \centering
  \includegraphics[width=0.35\textwidth]{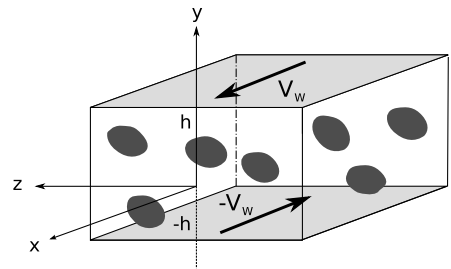}
  \caption{Sketch of the channel geometry and coordinate system adopted in this syudy.}
  \label{fig:sketch}
\end{figure}
We consider the flow of a suspension of deformable viscous hyperelastic particles in an incompressible Newtonian viscous fluid in a channel with moving walls, plane Couette geometry. The solid suspension have the same density $\rho$ as the fluid. The unstressed reference shape of the particle is a sphere of radius $r$. \figrefSC{fig:sketch} shows a sketch of the geometry and the Cartesian coordinate system, where $x$, $y$ and $z$ ($x_1$, $x_2$, and $x_3$) denote the streamwise, wall-normal and spanwise coordinates, while $u$, $v$ and $w$ ($u_1$, $u_2$, and $u_3$) denote the corresponding components of the velocity vector field. The lower and upper impermeable moving walls are located at $y=-h$ and $y=h$, and move in opposite direction with constant streamwise velocity $\pm V_w$.

The fluid and solid phase motion is governed by conservation of momentum and the incompressibility constraint:
\begin{subequations}
\label{eq:NS}
\begin{align}
\frac{\partial u_i^{\rm f}}{\partial t} + \frac{\partial u_i^{\rm f} u_j^{\rm f}}{\partial x_j} &= \frac{1}{\rho} \frac{\partial \sigma_{ij}^{\rm f}}{\partial x_j}, \\
\frac{\partial u_i^{\rm f}}{\partial x_i} &= 0, \\
\frac{\partial u_i^{\rm s}}{\partial t} + \frac{\partial u_i^{\rm s} u_j^{\rm s}}{\partial x_j} &= \frac{1}{\rho} \frac{\partial \sigma_{ij}^{\rm s}}{\partial x_j}, \\
\frac{\partial u_i^{\rm s}}{\partial x_i} &= 0,
\end{align}
\end{subequations}
where the suffixes $^{\rm f}$ and $^{\rm s}$ are used to indicate the fluid and solid phase. In the previous set of equations, $\rho$ is the density (assumed to be the same for the solid and fluid), and $\sigma_{ij}$ the Cauchy stress tensor. The kinematic and dynamic interactions between the fluid and solid phases are determined by enforcing the continuity of the velocity and traction force at the interface between the two phases
\begin{subequations}
\label{eq:bc}
\begin{align}
u_i^{\rm f} &= u_i^{\rm s}, \label{bc-v}\\
\sigmaf_{ij} n_j &= \sigmas_{ij} n_j \label{bc-sigma},
\end{align}
\end{subequations}
where $n_i$ denotes the normal vector at the interface. The fluid is assumed to be Newtonian
\begin{equation}
\label{eq:stress-f}
\sigmaf_{ij} = -p \delta_{ij} + 2 \muf D_{ij},
\end{equation}
where $\delta_{ij}$ is the Kronecker delta, $p$ is the pressure, $\muf$ the fluid dynamic viscosity, and $D_{ij}$ the strain rate tensor $(\partial u_i/\partial x_j + \partial u_j/\partial x_i)/2$. The solid is an incompressible viscous hyper-elastic material undergoing only the isochoric motion with constitutive equation
\begin{equation}
\label{eq:stress-s}
\sigmas_{ij} = -p \delta_{ij} + 2 \mus D_{ij} + G B_{ij},
\end{equation}
where $\mus$ is the solid dynamic viscosity, and the last term the hyper-elastic contribution modeled as a neo-Hookean material, thus satisfying the incompressible Mooney-Rivlin law, where $G$ is the modulus of transverse elasticity and $B_{ij}$ the left Cauchy-Green deformation tensor ($B_{ij} = F_{ik} F_{jk}$ where $F_{ij} = \partial x_i / \partial X_j$ is the deformation gradient, being $x$ and $X$ the current and reference coordinates \cite{bonet_wood_1997a}).

\begin{table*}
\centering
\setlength{\tabcolsep}{5pt}
\begin{tabular}{l|cccccccccc}
Case		&	$\color{brown}{+}$	&	$\color{brown}{\times}$	&	$\color{brown}{\ast}$	&	$\color{brown}{\boxdot}$	&	$\color{brown}{\bigtriangleup}$	&	$\color{blue}{+}$	&	$\color{blue}{\times}$	&	$\color{blue}{\ast}$	&	$\color{blue}{\boxdot}$	&	$\color{blue}{\bigtriangleup}$	\\
$\Phi$	&	$0.0016$		&	$0.0016$		&	$0.0016$		&	$0.0016$		&	$0.0016$		&	$0.11$	&	$0.11$	&	$0.11$	&	$0.11$	&	$0.11$	\\
$\Key$	&	$1$		&	$1$		&	$1$		&	$1$		&	$1$		&	$1$	&	$1$	&	$1$	&	$1$	&	$1$\\
$\Ca$	&	$0.02$		&	$0.1$		&	$0.2$		&	$0.4$		&	$2$		&	$0.02$	&	$0.1$	&	$0.2$	&	$0.4$	&	$2$\\ \hline
Case		&	$\color{orange}{+}$	&	$\color{orange}{\times}$	&	$\color{orange}{\ast}$	&	$\color{orange}{\boxdot}$	&	$\color{orange}{\bigtriangleup}$	&	$\color{red}{+}$	&	$\color{red}{\times}$	&	$\color{red}{\ast}$	&	$\color{red}{\boxdot}$	&	$\color{red}{\bigtriangleup}$\\
$\Phi$	&	$0.22$	&	$0.22$	&	$0.22$	&	$0.22$	&	$0.22$	&	$0.33$	&	$0.33$	&	$0.33$	&	$0.33$	&	$0.33$	\\
$\Key$	&	$1$		&	$1$		&	$1$		&	$1$		&	$1$		&	$1$	&	$1$	&	$1$	&	$1$	&	$1$\\
$\Ca$	&	$0.02$		&	$0.1$		&	$0.2$		&	$0.4$		&	$2$		&	$0.02$	&	$0.1$	&	$0.2$	&	$0.4$	&	$2$\\ \hline
Case		&	$\color{blue}{\Diamond}$	&	$\color{blue}{\bigcirc}$	&	$\color{blue}{\bigtriangledown}$	&&&&&&&	\\
$\Phi$	&	$0.11$	&	$0.11$	&	$0.11$	&&&&&&&	\\
$\Key$	&	$0.01$		&	$0.1$		&	$10$		&&&&&&&	\\
$\Ca$	&	$0.2$		&	$0.2$		&	$0.2$	&&&&&&&
\end{tabular}
\caption{Summary of the DNSs performed, all at a fixed particle Reynolds number, $\Rey = \rho \dot{\gamma} r^2/\muf =0.1$.}
\label{tab:cases}
\end{table*}

To numerically solve the fluid-structure interaction problem at hand, we adopt the so called one-continuum formulation \cite{tryggvason_sussman_hussaini_2007a}, where only one set of equations is solved over the whole domain. This is obtained by introducing a monolithic velocity vector field valid everywhere, found by applying the volume averaging procedure \cite{takeuchi_yuki_ueyama_kajishima_2010a, quintard_whitaker_1994b}. Thus, we can write the Cauchy stress tensor $\sigma_{ij}$ in a mixture form, similarly to the Volume of Fluid \citep{hirt_nichols_1981a} and Level Set \citep{sussman_smereka_osher_1994a, chang_hou_merriman_osher_1996a} methods commonly used to simulate multiphase flows:
\begin{equation}
\label{eq:phi-stress}
\sigma_{ij} = \left( 1 - \phis \right) \sigmaf_{ij} + \phis \sigmas_{ij},
\end{equation}
where $\phis$ is a local phase indicator based on the local solid volume fraction. Thus, at each point of the domain the fluid and solid phases are distinguished by $\phis$, which is equal to $0$ in the fluid, $1$ in the solid, and between $0$ and $1$ in the interface cells.  The set of equations can be closed in a purely Eulerian manner by introducing a transport equation for the volume fraction $\phis$
\begin{equation}
\label{eq:PHI-adv}
\frac{\partial \phis}{\partial t} + \frac{\partial u_k \phis}{\partial x_k} = 0,
\end{equation}
and updating the left Cauchy-Green deformation tensor components with the following transport equation:
\begin{equation}
\label{eq:B-adv}
\frac{\partial B_{ij}}{\partial t} + \frac{\partial u_k B_{ij}}{\partial x_k} = B_{kj}\frac{\partial u_i}{\partial x_k} + B_{ik}\frac{\partial u_j}{\partial x_k},
\end{equation}
expressing the fact that the upper convected derivative of the left Cauchy-Green deformation tensor is identically zero \cite{bonet_wood_1997a}. 

The equations are solved numerically: the time integration is based on an explicit fractional-step method \cite{kim_moin_1985a}, where all the terms are advanced with the third order Runge-Kutta scheme, except the solid hyper-elastic contribution which is advanced with the Crank-Nicolson scheme \cite{min_yoo_choi_2001a}. The governing differential equations are solved on a staggered grid using a second order central finite-difference scheme, except for the advection terms in \equref{eq:PHI-adv} and \equref{eq:B-adv} where the fifth-order WENO scheme is applied. The code has been extensively validated, and more details on the numerical scheme and validation campaign are reported in Ref.~\cite{rosti_brandt_2017a}. Note that, no special artifact is used to avoid or model the particle-wall and particle-particle interaction, because the hydrodynamic repulsion is brought by the soft lubrication effect due to the geometry change via the particle deformation \cite{skotheim_mahadevan_2005a}. More details on the numerical method can be found in Ref.~\cite{sugiyama_ii_takeuchi_takagi_matsumoto_2011a}.

\subsection{Numerical setup}
We consider the Couette flow of a Newtonian fluid laden with hyper-elastic deformable spheres. The Reynolds number of the simulation is fixed to $\Rey = \rho \dot{\gamma} r^2/\muf = 0.1$, where $\dot{\gamma}$ is the reference shear rate, so that we can consider inertial effects negligible. The total solid volume fraction of the suspension $\Phi$ is defined as the volume average of the local volume fraction $\phi$, \ie $\Phi=\bra{\bra{\phi}}$. Hereafter, the double $\bra{\bra{\cdot}}$ indicates the time and volume average while the single $\bra{\cdot}$ the average in time and in the homogeneous $x$ and $z$ directions. Four values of total volume fraction $\Phi \approx 0.0016$, $0.11$, $0.22$, and $0.33$ are considered, together with five values of elastic moduli $G$, resulting in the capillary numbers $\Ca = \muf \dot{\gamma}/G = 0.02$, $0.1$, $0.2$, $0.4$, and $2$. For all the previous cases the solid viscosity is set equal to the fluid viscosity, \ie $\Key=\mus/\muf=1$. To study the effect of the solid viscosity, we run three additional simulations for the case with $\Phi=0.11$ and $\Ca=0.2$ with the following solid/fluid viscosity ratio $\Key=\mus/\muf=0.01$, $0.1$, and $10$. The full set of simulations analysed is reported in \tabref{tab:cases}, together with the color scheme and symbols used in the figures throughout the manuscript. Note that, the range of the viscoelsatic parameters ($\Ca$ and $\Key$), as well as the volume fraction $\Phi$, are similar to that of many previous works that can be found in the literature \cite{davino_greco_hulsen_maffettone_2013a, picano_breugem_mitra_brandt_2013a, villone_hulsen_anderson_maffettone_2014a, matsunaga_imai_yamaguchi_ishikawa_2016a, villone_greco_hulsen_maffettone_2016a}. The numerical domain is a rectangular box of size $16r \times 10r \times 16r$ in the $x$, $y$, and $z$ directions, discretised on a Cartesian uniform mesh with $16$ grid points per sphere radius $r$. No-slip boundary conditions are imposed on the solid walls, while periodic boundary conditions are enforced in the homogeneous $x$ and $z$ directions. All the simulations are started from a stationary flow with a random distribution of the particles across the domain. The grid independence of the results has been verified by simulating the case with $\Phi=0.22$ and $\Ca=0.2$ with double grid points in each direction, resulting in a difference in the effective viscosity lower than $0.5\%$. The effect of the size of the domain on the results (\eg confinement effects) has not been studied in the present work, and the domain size was chosen the same as in a previous study \cite{picano_breugem_mitra_brandt_2013a}; the interested reader is referred to Ref.~\cite{fornari_brandt_chaudhuri_lopez_mitra_picano_2016a} for more details on confined suspensions in a similar geometry. Note that, similarly to the box size, also the general set-up and most of parameters used in this study are chosen as in Ref.~\cite{picano_breugem_mitra_brandt_2013a} where suspensions of rigid spheres are examined to ease comparisons.

\begin{figure}[t]
  \centering
  \input{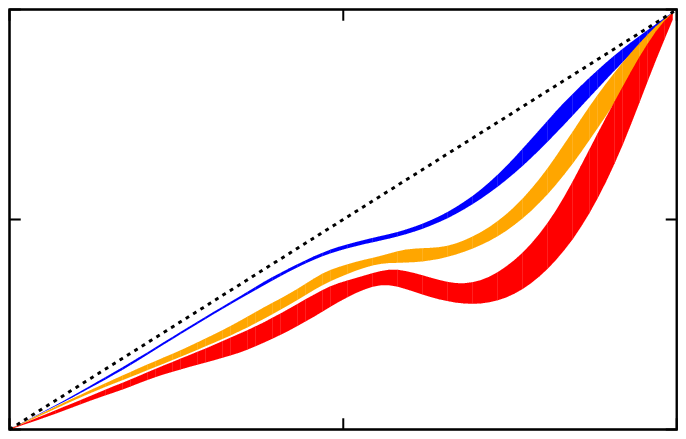}\\
  \vspace{0.55cm}
  \input{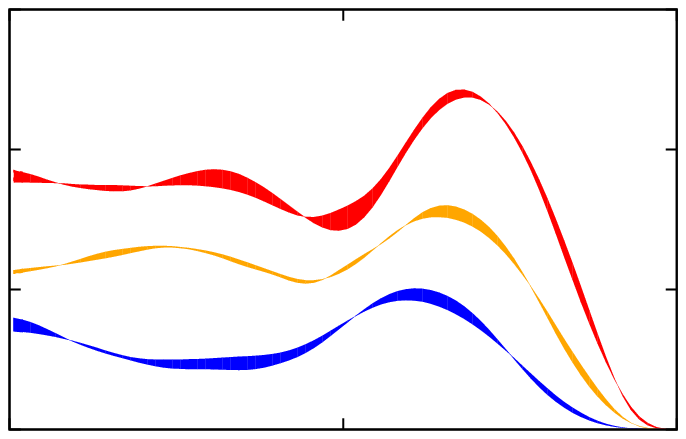}\\
  \vspace{0.55cm}
  \caption{(top) Mean fluid streamwise velocity profile $\vf$, normalized with the wall velocity $V_w$, and (bottom) mean local volume fraction distribution $\phis$ along the wall-normal direction $y$.}
  \label{fig:velVof}
\end{figure}
\begin{figure}[t]
  \centering
  \input{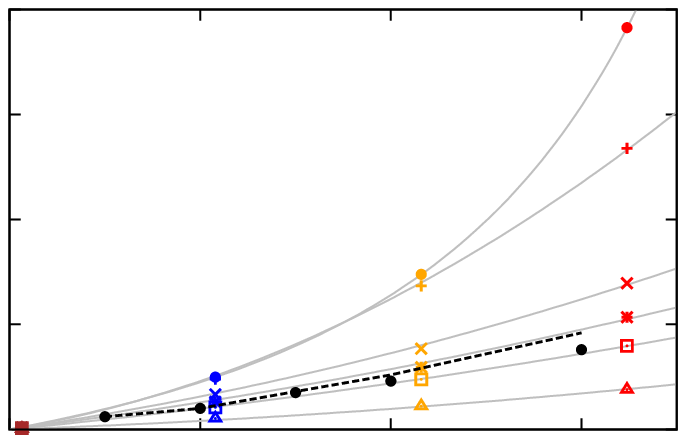}\\
  \vspace{0.55cm}  
  \input{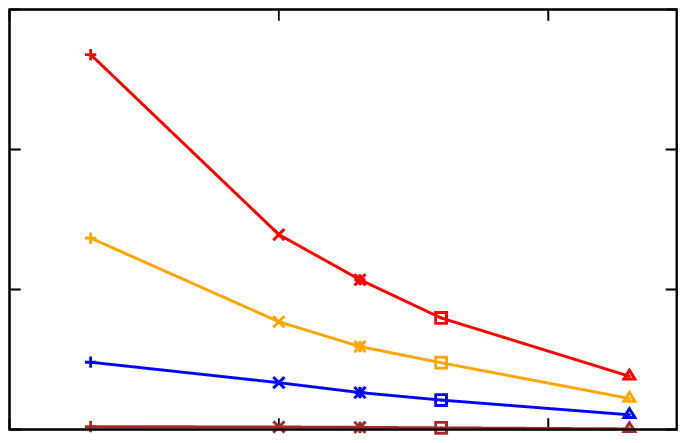}\\
  \vspace{0.55cm}
  \caption{The effective viscosity $\mu/\muf$ as a function of (top) the volume fraction $\Phi$ and of (bottom) the Capillary number $\Ca$, for several different values of volume fractions $\Phi \approx 0.0016$ (brown), $0.11$ (blue), $0.22$ (orange), and $0.33$ (red) and of Capillary numbers $\Ca=0.02~(+)$, $0.1~(\times)$, $0.2~(\ast)$, $0.4~(\boxdot)$, and $2~(\bigtriangleup)$.  All the cases have $\Key=1$. For comparison we also plot the same data for rigid particles~\cite{picano_breugem_mitra_brandt_2013a} ($\Ca=0$ (color circles), the results for drops with $Ca=0.15$ from Ref.~\cite{srivastava_malipeddi_sarkar_2016a} (black circles) and Pal's empirical relation \cite{pal_2003a} (black dashed line).}
  \label{fig:viscFG}
\end{figure}

\section{Results} \label{sec:result}
\begin{figure*}[]
  \centering
  \includegraphics[width=0.322\textwidth]{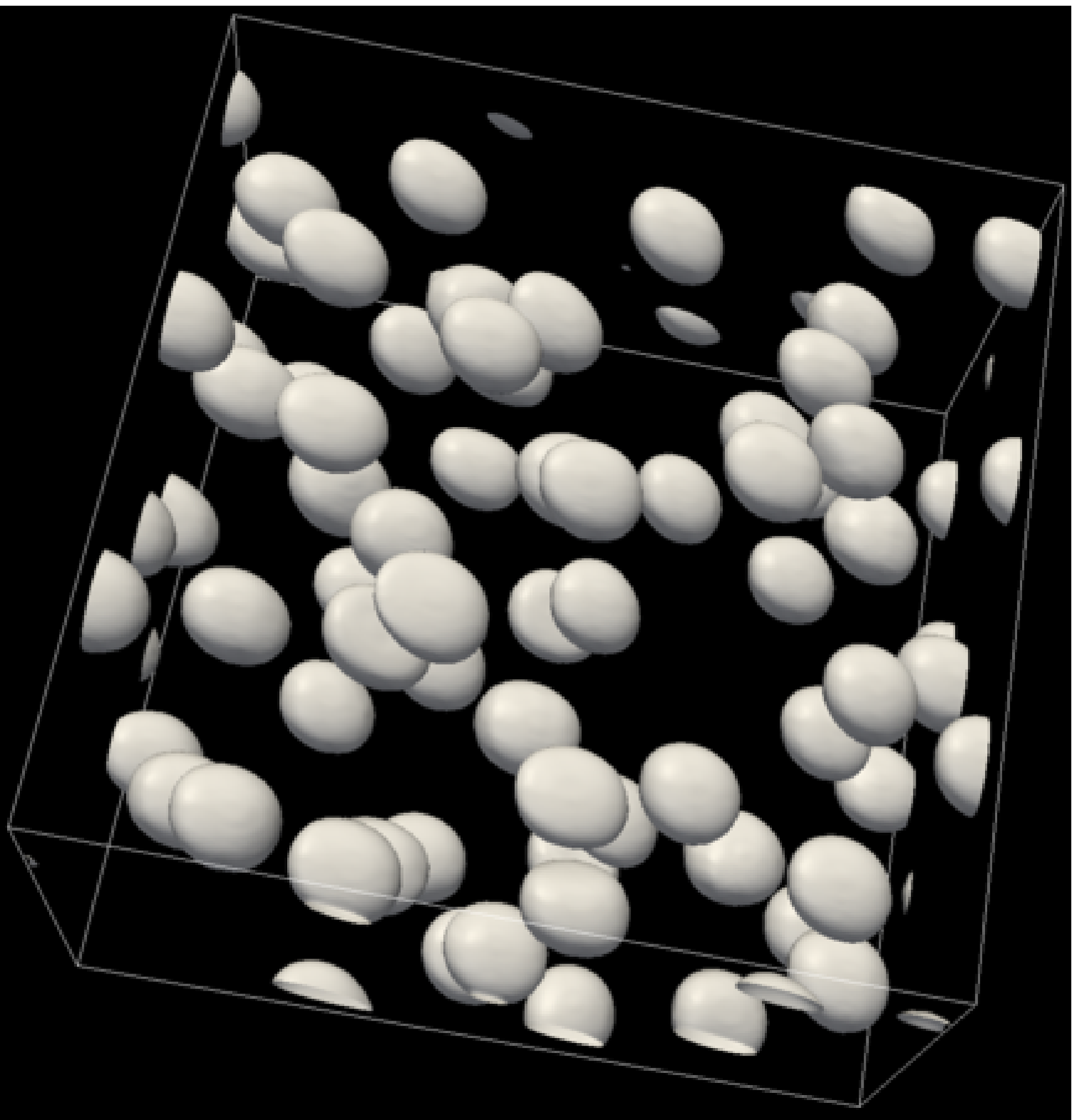}
  \includegraphics[width=0.322\textwidth]{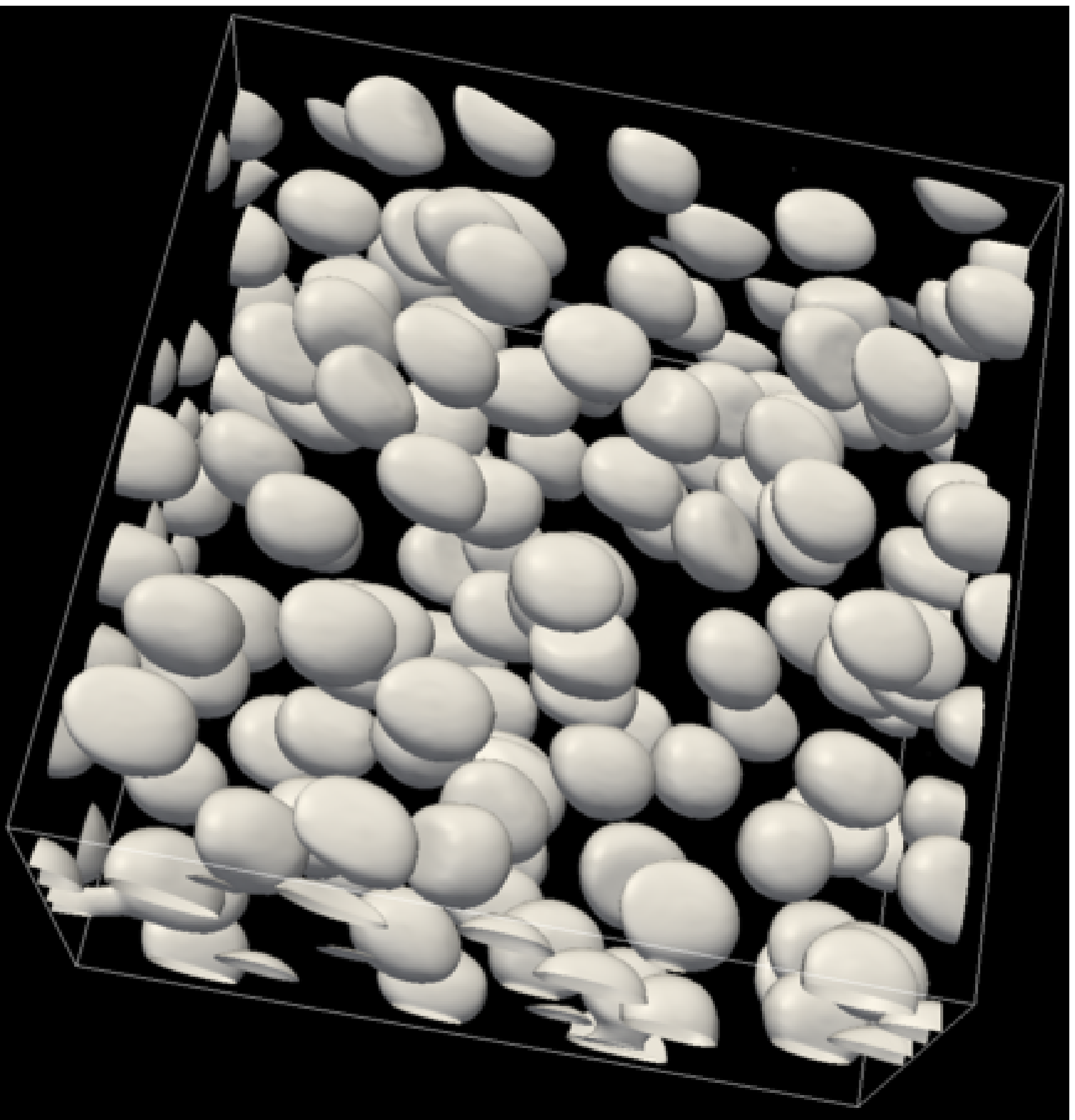}
  \includegraphics[width=0.322\textwidth]{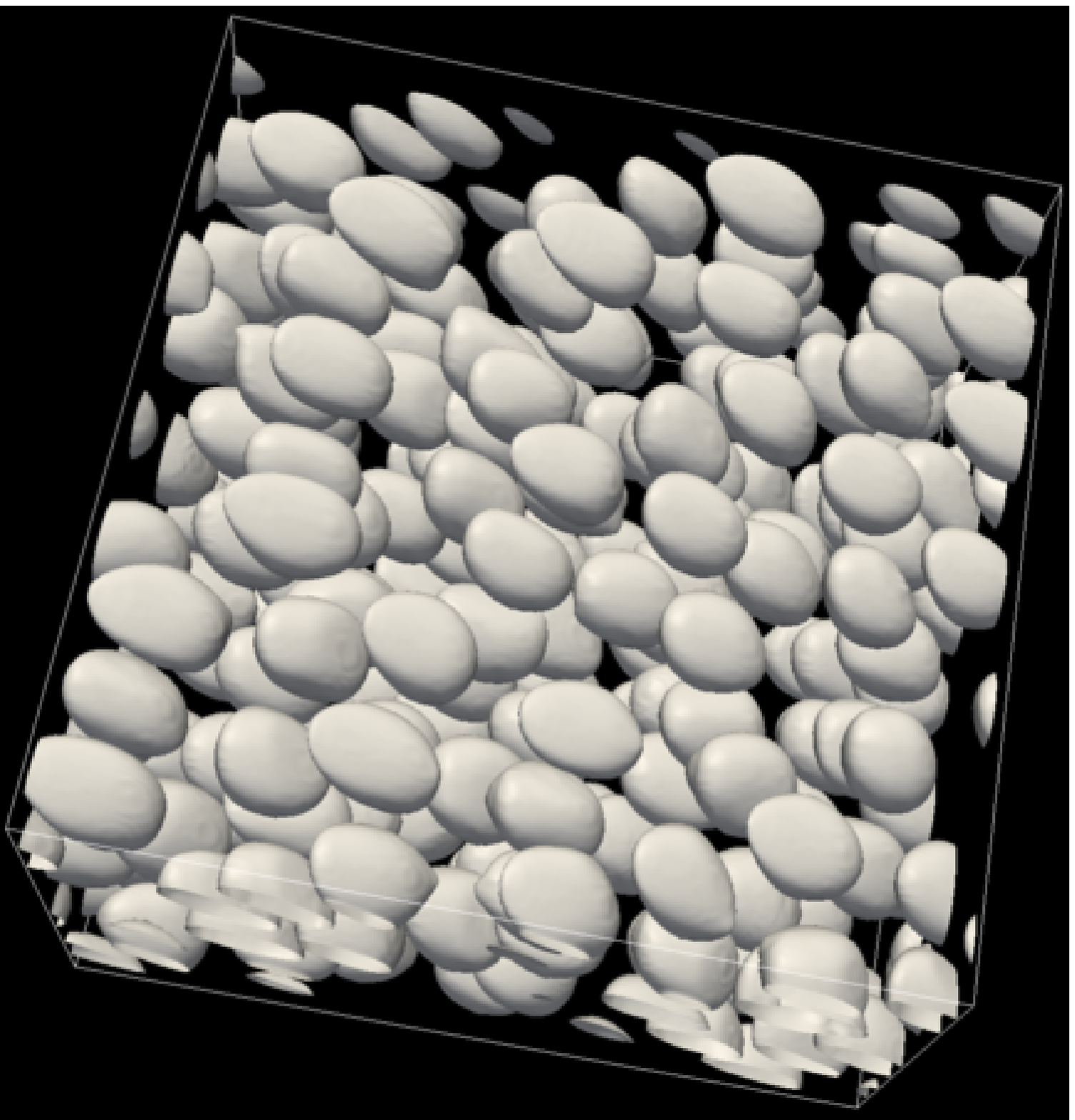}\\ \vspace{0.07cm}
  \includegraphics[width=0.19\textwidth]{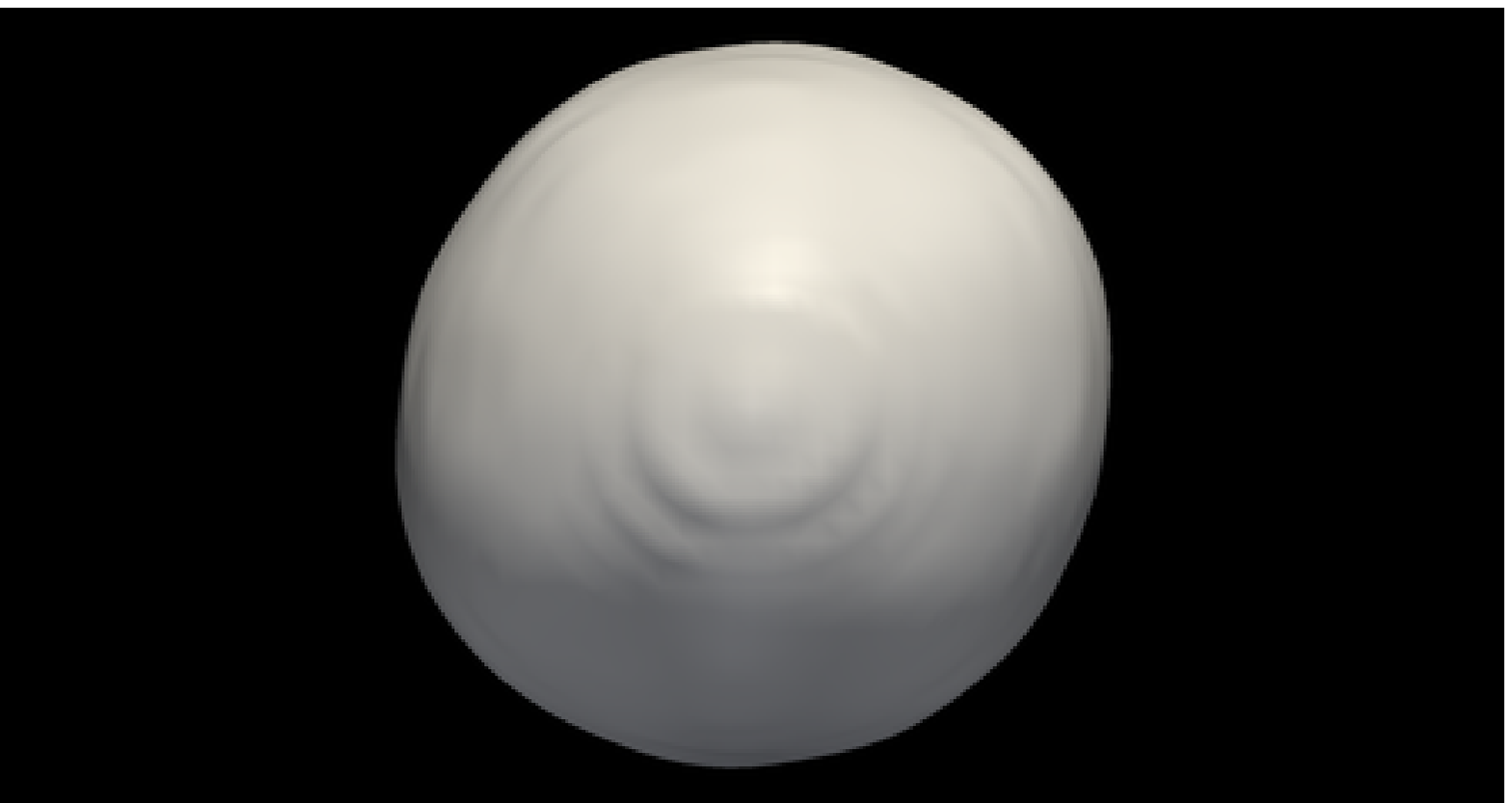}
  \includegraphics[width=0.19\textwidth]{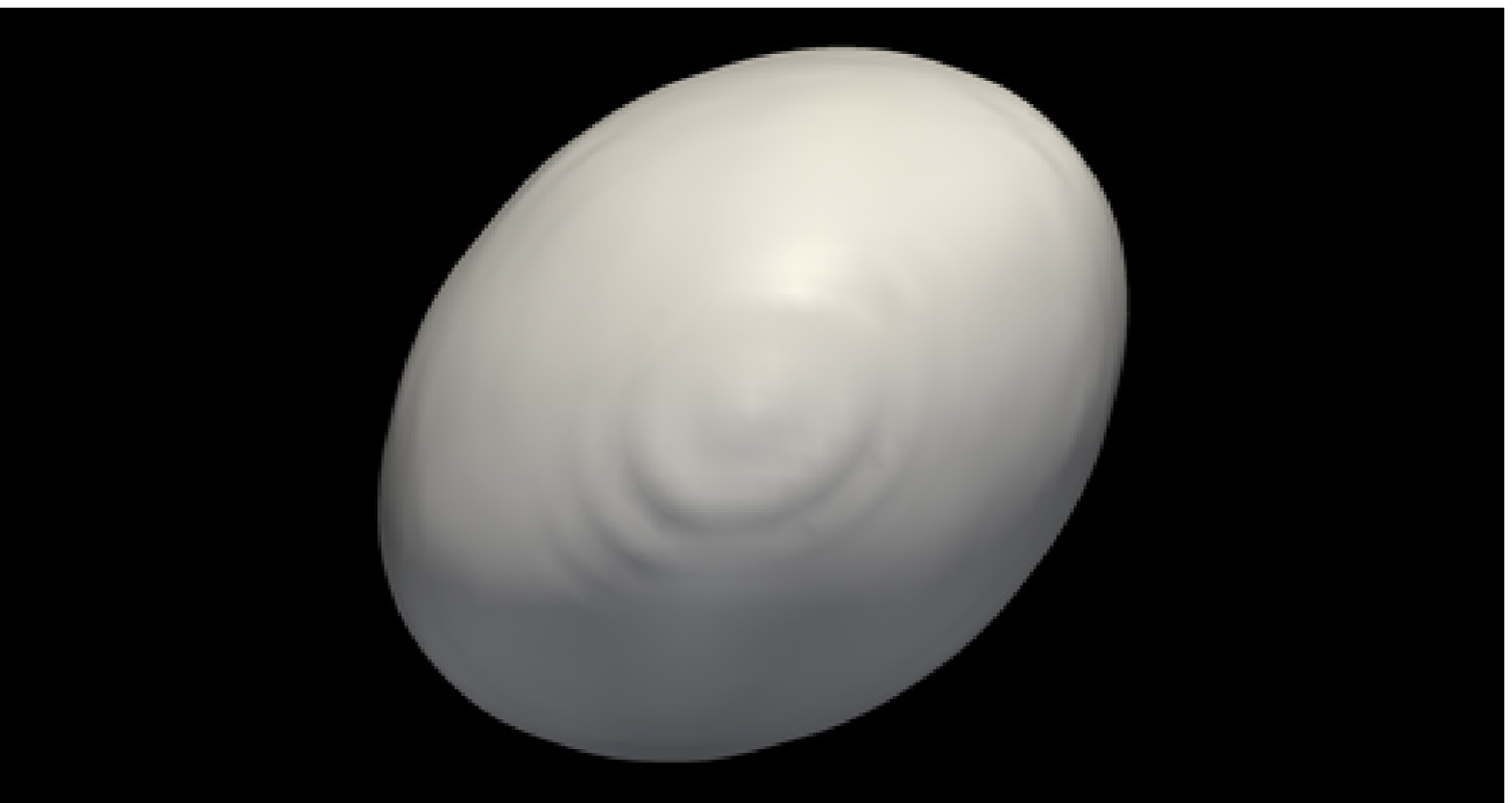}
  \includegraphics[width=0.19\textwidth]{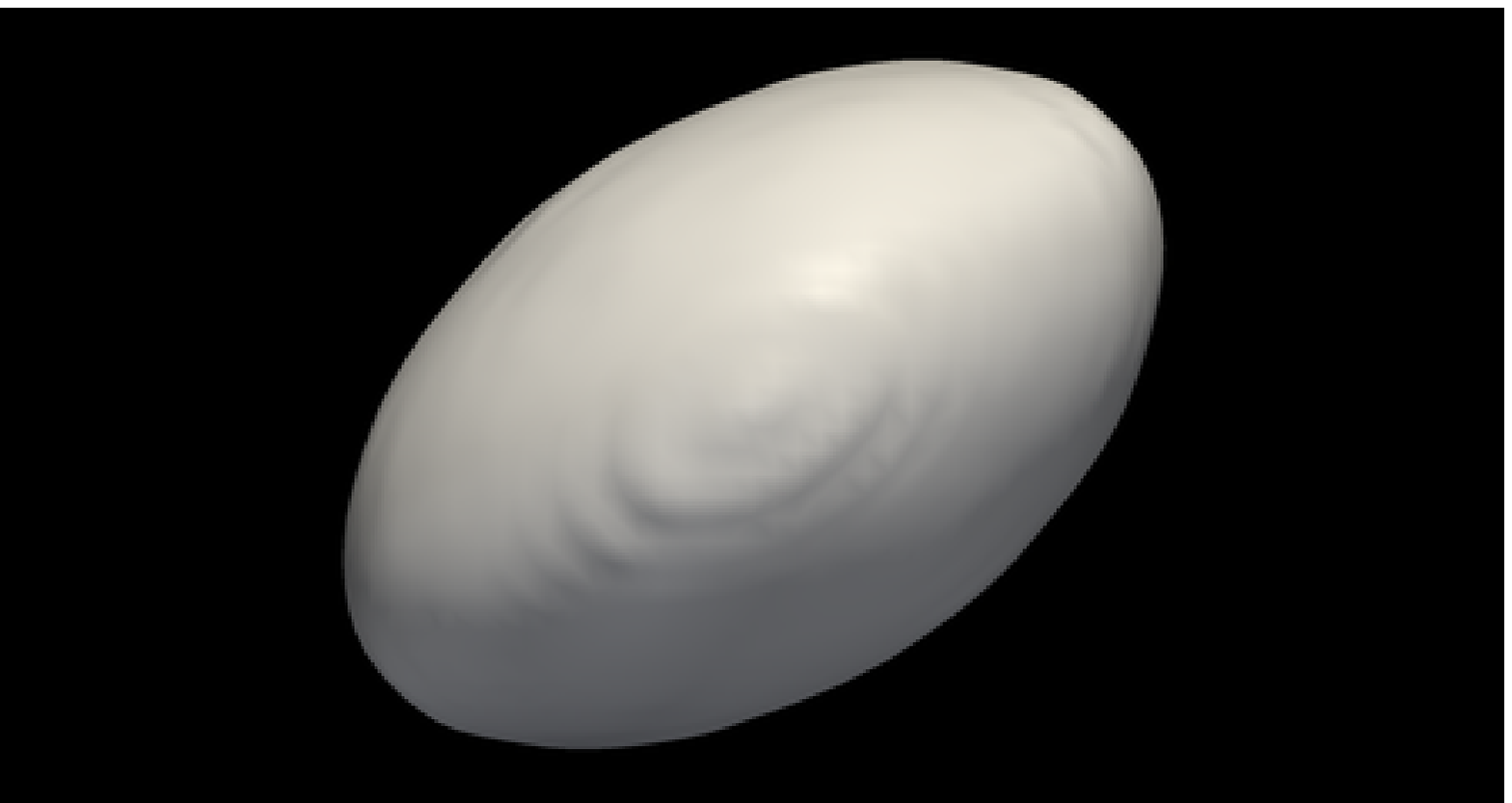}
  \includegraphics[width=0.19\textwidth]{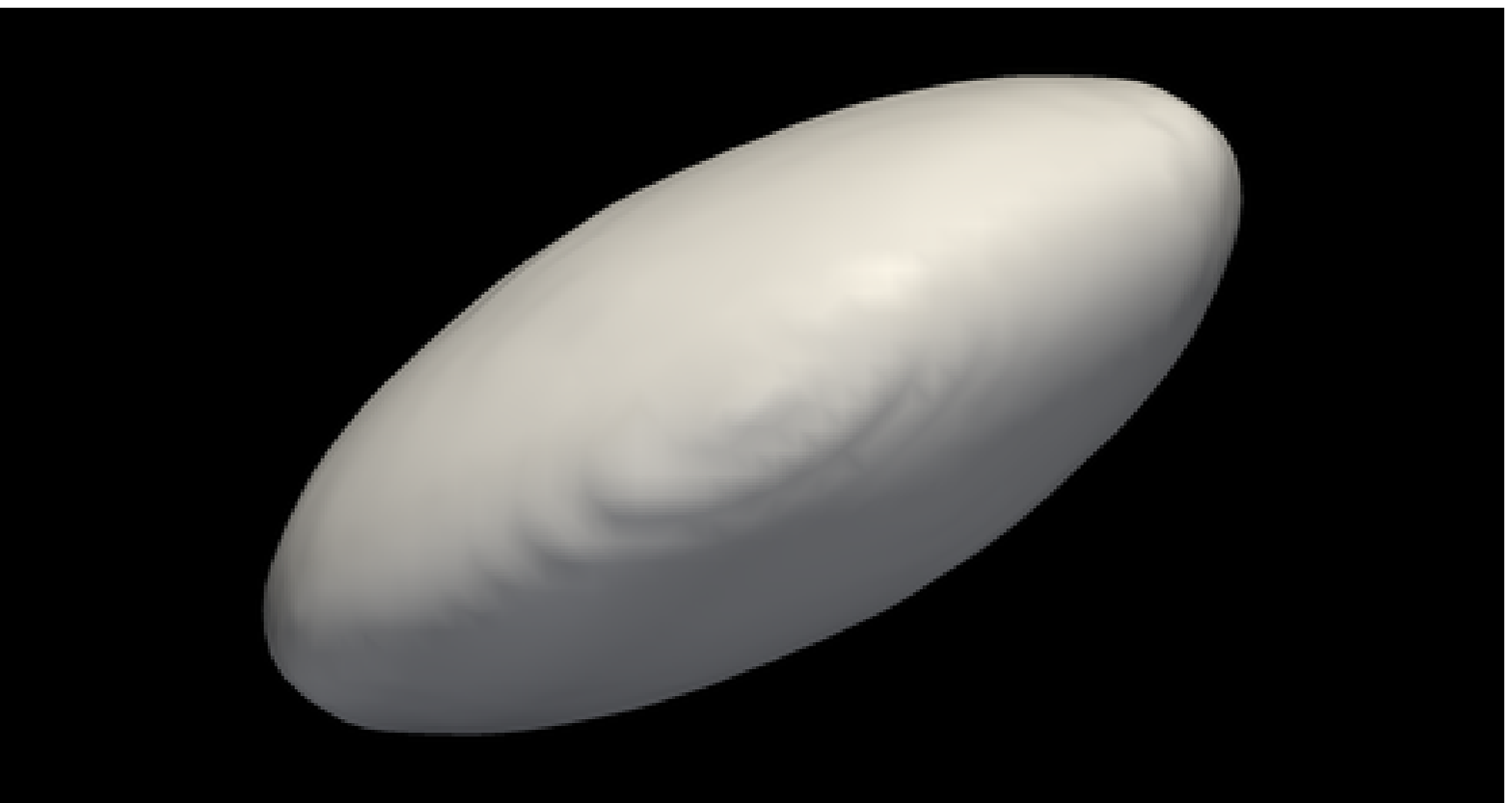}
  \includegraphics[width=0.19\textwidth]{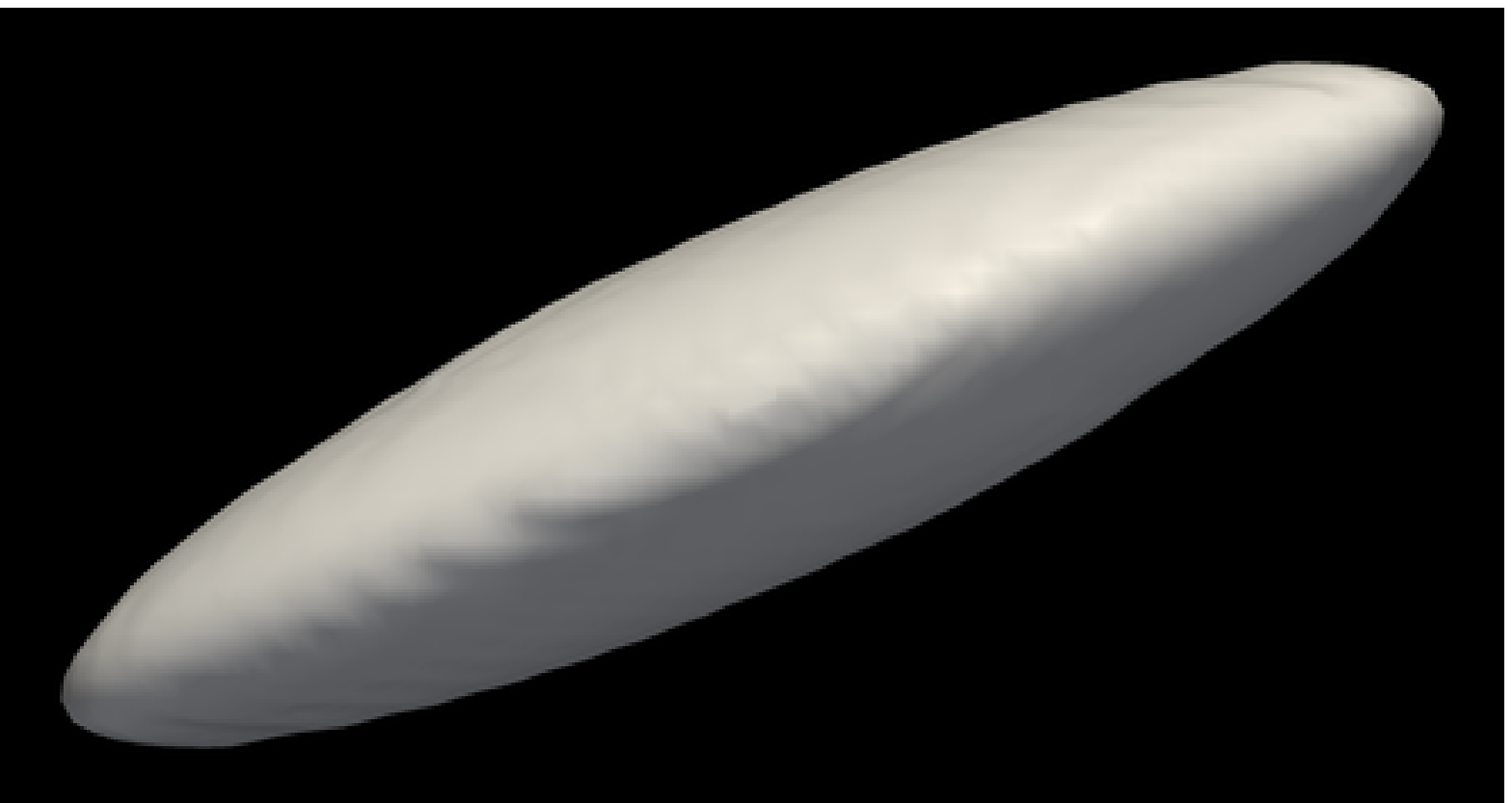}
  \caption{(top) Instantaneous shape of the deformed particles at $\Ca=0.2$ for three different volume fractions: $\Phi=0.11$, $0.22$, and $0.33$. (bottom) Shape of the particle in the dilute case, $\Phi=0.0016$, for increasing Capillary numbers $\Ca=0.02$, $0.1$, $0.2$, $0.4$ and $2$.}
  \label{fig:shape}
\end{figure*}
We start the analysis of the suspension of deformable particles by showing the mean streamwise velocity profile $\vf$ and the solid concentration $\phis$ in \figrefS{fig:velVof}, where the shaded area represents the spread of the data due to the different Capillary numbers. The mean streamwise velocity equals $V_w$ at the wall due to the no-slip condition with the moving wall, and is null at the center line for symmetry. The velocity profile, a straight line for a Newtonian fluid, is clearly different due to the presence of the suspended particles. In particular, the velocity decreases faster than the Newtonian case close to wall, \ie the wall-normal derivative of the velocity profile at the wall increases, it shows a local minimum around $y\approx0.75h$ and then goes smoothly to zero; these differences are enhanced for high values of volume fractions and Capillary numbers, especially in the near wall region and around the local minimum. The shape of the velocity profile is strongly related to the mean distribution of the particles across the channel, as shown by the wall-normal profiles of the local solid volume fraction in the bottom panel of the same figure. The particles have a non uniform distribution in the $y$ direction, with a clear layer close to the wall as shown by the peak in the concentration around $y\approx0.75h$. This is due to the excluded volume effects at the wall \cite{yeo_maxey_2010a, picano_breugem_mitra_brandt_2013a}, and corresponds to the position of the local minimum of velocity. Also, we note that the location of maximum particle concentration moves towards the wall for increasing volume fractions.

The wall-normal derivative of the streamwise velocity at the wall can be used to estimate the effective viscosity of the non-Newtonian fluid made by the suspension of particles in the Newtonian fluid. Indeed, we define the effective suspension viscosity $\mu$, normalized by the fluid one $\muf$, as follows
\begin{equation} \label{eq:visc}
\frac{\mu}{\muf}=\frac{\bra{\sigma_{12}^{\rm w}}}{\muf \dot{\gamma}}.
\end{equation}
The effective viscosity as a function of the total volume fraction $\Phi$ and of the Capillary number $\Ca$ is shown in \figrefS{fig:viscFG}. Here, the solid colour circles represent the limiting case of completely rigid particles ($\Ca=0$), taken from the simulations in Ref.~\cite{picano_breugem_mitra_brandt_2013a}, while the black circles are the results for drops with $Ca=0.15$ from Ref.~\cite{srivastava_malipeddi_sarkar_2016a} and the dashed line Pal's empirical relation \cite{pal_2003a}. We observe that the effective viscosity is a monotonic non-linear function of both variables, and in particular, it increases with the volume fraction $\Phi$ and decreases with the Capillary number $\Ca$. All the deformable cases have lower effective  viscosity then the rigid ones at the same volume fraction, and the difference is enhanced for the higher values of the Capillary number. Note also that the growth rate of the effective viscosity is reducing for increasing Capillary numbers, and that it appears to be almost linear for the highest $Ca$ under consideration, something usually associated to suspensions of monodisperse particles ($\Phi \rightarrow 0$) \cite{einstein_1956a}. As clearly shown in the figure, the limit for $\Ca \rightarrow 0$ is the rigid particle behavior, while for $\Ca \rightarrow \infty$ the suspension viscosity approaches that of the fluid, \ie $\mu/\muf \rightarrow 1$. In the top panel of \figrefS{fig:viscFG}, we display with grey lines a fit to our data. The data for rigid spheres collapse well onto the Eilers fit; this is an experimental fit for non-Brownian particles at zero Reynolds number, valid also for high volume fractions \cite{zarraga_hill_leighton-jr_2000a, singh_nott_2003a, kulkarni_morris_2008a}, reading
\begin{equation} \label{eq:Eilers}
\dfrac{\mu}{\mu^f} = \left[ 1 + B_E \dfrac{\Phi}{1-\Phi/\Phi_m}\right]^2,
\end{equation}
where $\Phi_m=0.58 - 0.63$ is the geometrical maximum packing, and $B_E =1.25 - 1.7$ a coefficient. Here we use $\Phi_m=0.6$ and $B_E=1.7$. The simulations pertaining suspensions of deformable particles are fitted with by an expression formally similar to the Batchelor and Green relation \cite{batchelor_green_1972a} , \ie a second order extension of the Einstein's formula \cite{einstein_1956a},
\begin{equation} \label{eq:batchelor}
\dfrac{\mu}{\muf}=1+\left[\mu\right]\Phi+B_B \Phi^2,
\end{equation}
where $\left[\mu\right]$ is the intrinsic viscosity equal to $5/2$ for rigid dilute particles and $B_B$ is a coefficient equal to $7.6$ for non-Brownian spheres. For the cases of deformable particles studied here, we use the intrinsic viscosity computed from simulations with a single sphere ($\Phi \approx 0.0016$ for the domain considered here), while $B_B$ is kept as a fitting parameter. This second-order relation is usually inaccurate for $\Phi \gtrsim 0.15$, when the viscosity starts increasing faster than a second order polynomial \cite{stickel_powell_2005a}. However, we show here that it remains valid for values of the volume fraction up to $\Phi=0.33$ in the case of deformable particles, provided the intrinsic viscosity $\left[\mu\right]$ is modified to take into account the change of shape. In other words, in the case of deformable particles, both $\left[\mu\right]$ and $\Phi$ are function of the Capillary number $\Ca$. The fitting coefficients obtained by the method of least squares are reported in \tabref{tab:fit}, together with the intrinsic viscosity extracted from our simulations. This is computed as $\left[\mu\right] \approx \left( \mu - \muf \right)/ \left( \muf \Phi \right)$.
\begin{table}
\centering
\setlength{\tabcolsep}{5pt}
\begin{tabular}{l|cc}
$\Ca$	&	$\left[ \mu \right]$	&	$B_B$ 			\\ \hline
$0.02$	&	$2.988$					&	$16.15$	\\
$0.1$	&	$2.633$					&	$4.998$	\\
$0.2$	&	$2.274$					&	$2.988$	\\
$0.4$	&	$1.779$					&	$2.069$	\\
$2$		&	$0.641$					&	$1.677$
\end{tabular}
\caption{The fitting parameter $B_B$ in \equref{eq:batchelor} used for the curves in \figrefS{fig:viscFG} and the intrinsic viscosity $\left[\mu\right]$ computed from our simulations.}
\label{tab:fit}
\end{table}

\begin{figure}[t]
  \centering
  \input{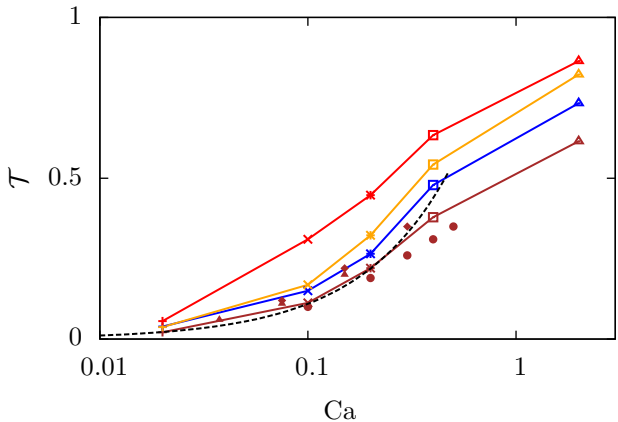}\\
  \vspace{0.55cm}
  \caption{Taylor deformation parameter $\mathcal{T}$, defined in \equref{eq:Taylor}, as a function of the Capillary number $\Ca$ for different values of volume fraction $\Phi$. The symbol and color scheme is the same as in \figrefS{fig:viscFG}. The brown symbols are numerical results from the literature for a single sphere in a box. In particular, the triangles and rhombus are the results from Refs.~\cite{pozrikidis_1995a,eggleton_popel_1998a} which were calculated for $\Phi\approx0.06$, while the circles are the $2$D simulation from Ref.~\cite{ii_sugiyama_takeuchi_takagi_matsumoto_2011a} with $\Phi\approx0.05$. The black dashed line shows the result of the perturbative calculation of Ref~\cite{choi_schowalter_1975a} expected to hold for small $\Ca$ and $\Phi$.}
  \label{fig:def}
\end{figure}

\begin{figure}[t]
  \centering
  \input{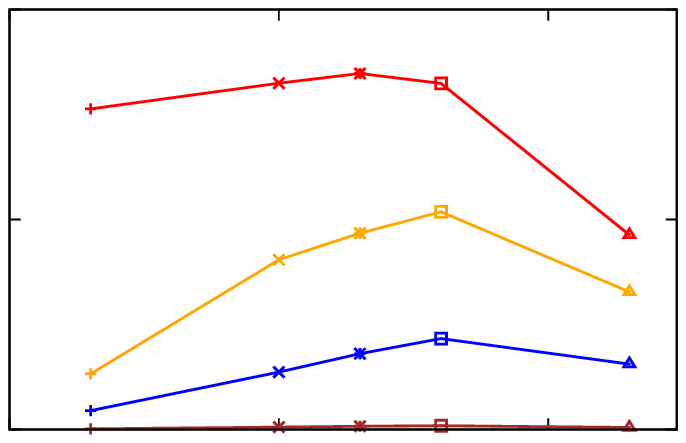}\\
  \vspace{0.55cm}  
  \input{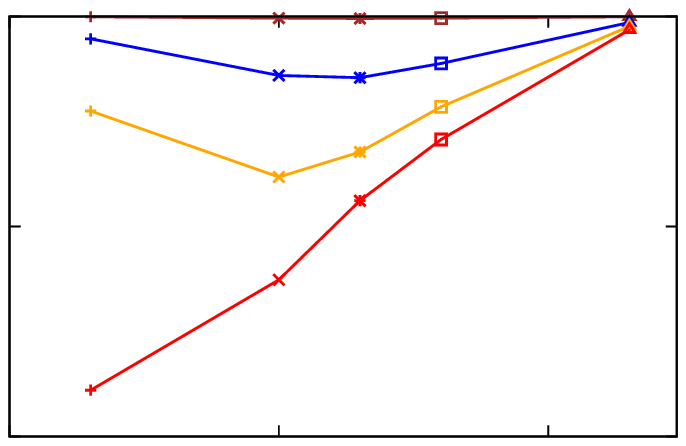}\\
  \vspace{0.55cm}
  \caption{(top) First and (bottom) second normal stress difference, $\NSI$ and $\NSII$, as a function of the Capillary number $\Ca$. The color and symbol scheme is the same as in the previous figures. The normal stresses are normalized by $\muf\dot{\gamma}$.}
  \label{fig:normStress}
\end{figure}

Next, we study the particle deformation. \figrefSC{fig:shape} shows instantaneous particle configurations: the top row represents the cases at $\Ca=0.2$ and increasing $\Phi$, \ie $0.11$, $0.22$ and $0.33$, while the bottom row is for the dilute suspension $\Phi=0.0016$ and increasing $\Ca$, \ie $0.02$, $0.1$, $0.2$, $0.4$ and $2$. We observe that the shape is strongly affected by $\Ca$ and weakly by $\Phi$. In particular, the particles, originally spheres,  progressively become elongated ellipsoids as $\Ca$ increases. To quantify this effect, we therefore evaluate the deformation by means of the so-called Taylor parameter
\begin{equation} \label{eq:Taylor}
\mathcal{T}=\frac{a-b}{a+b},
\end{equation}
where $a$ and $b$ are the semi-major and semi-minor axis of the inscribed ellipse passing through the particle center in the $x$-$y$ plane, and report its dependency on $Ca$ in \figrefS{fig:def}. Note that, the parameter $\mathcal{T}$ is averaged over all the particles and in time. In the figure results from the literature, obtained in the limit of $\Phi \rightarrow 0$, are also reported as comparison. We note that the Taylor parameter increases with both the Capillary number and the volume fraction, the former being more effective than the latter. The trend is the same as obtained for dilute systems, as shown by the brown curve in \figrefS{fig:def}, pertaining our results for a single particle, and by the symbols indicating results in the literature  \cite{pozrikidis_1995a,eggleton_popel_1998a,ii_sugiyama_takeuchi_takagi_matsumoto_2011a}.

To fully characterize the non-Newtonian suspension behavior, we report in \figrefSC{fig:normStress} the first and second normal stress difference, $\NSI$ and $\NSII$, as a function of the Capillary number $\Ca$ for all the volume fractions under investigations. These are a measure of the viscoelasticity of the flow and are defined as
\begin{subequations}
\label{eq:normStress}
\begin{align}
\NSI  &=\bra{\bra{\sigma_{11}-\sigma_{22}}}, \\
\NSII &=\bra{\bra{\sigma_{22}-\sigma_{33}}}.
\end{align}
\end{subequations}
In the Eulerian framework adopted here, they can be easily computed from \equref{eq:phi-stress}. The first normal stress difference $\NSI$ is positive whereas the second one $\NSII$ is negative, with $\vert \NSI/\NSII \vert>1$. This behaviour is typical of most viscoelastic fluids, and corresponds to the fluid forcing the walls apart \cite{mewis_wagner_2012a}; it is similar to the one observed for suspensions of capsules \cite{matsunaga_imai_yamaguchi_ishikawa_2016a}, filaments \cite{wu_aidun_2010b} and polymers.  We observe that an increase in volume fraction leads to an increase of the first normal stress difference $\NSI$ and a decrease of the second one $\NSII$ (increase in the absolute value). Both the normal stress differences are non-monotonic with $\Ca$, showing a maximum and minimum which moves at lower $\Ca$ for increasing volume fractions, which is similar to what discussed in Ref.~\cite{wu_aidun_2010b} for a filament suspension. Our results are in agreement with the data from the simulations of elastic capsules in Ref.~\cite{matsunaga_imai_yamaguchi_ishikawa_2016a}, where, however, the non-monotonic behavior of the normal stress difference is not appearing owing to a more limited range of Capillary number considered.

\begin{figure}[]
  \centering
  \input{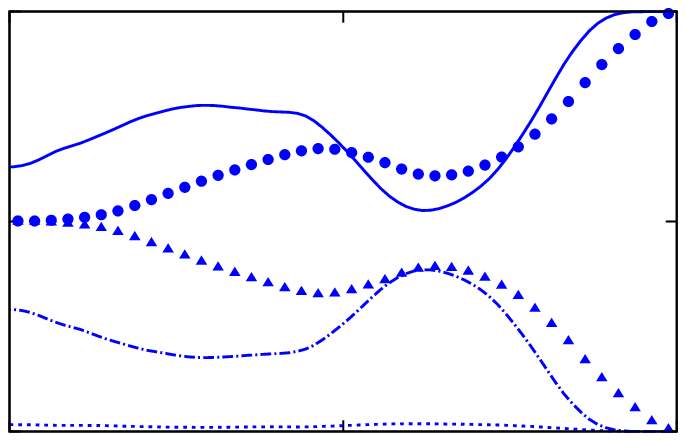}\\
  \vspace{0.55cm}
  \input{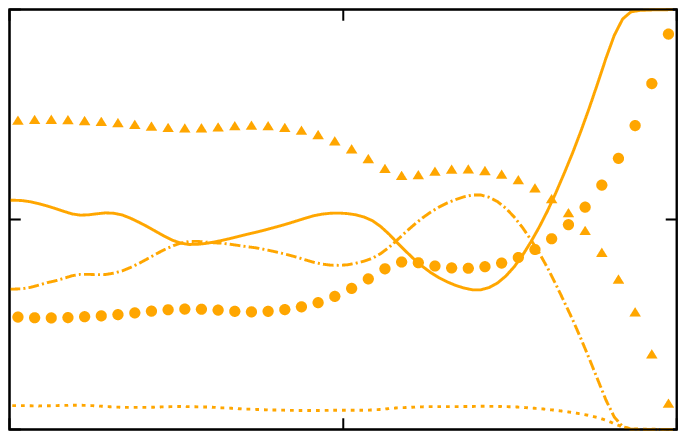}\\
  \vspace{0.55cm}  
  \input{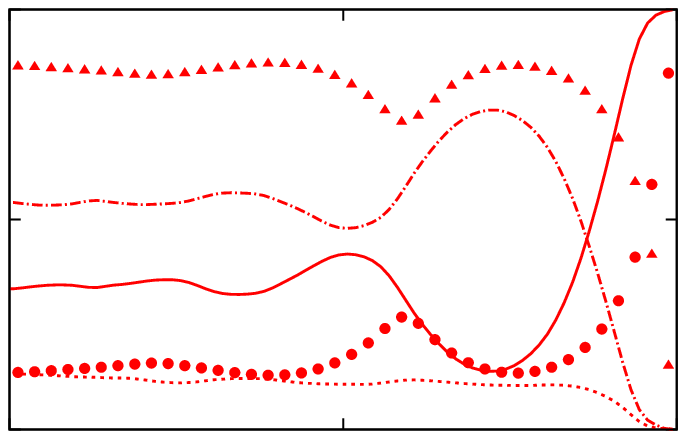}\\
  \vspace{0.55cm}
  \caption{Decomposition of the total shear stress $\bra{\sigma_{12}}$ in its contributions, normalized by the total wall value $\bra{\sigma_{12}^{\rm w}}$. The color scheme is the same as in the previous figure, with blue, orange, and red curves indicating the different volume fractions $\Phi=0.11$ (top panel), $0.22$ (middle panel) and $0.33$ (bottom panel). In each plot, the symbols are used for the rigid case ($\Ca=0$), the circles $\bullet$ and triangles $\blacktriangle$ being the fluid viscous and particle stresses, and the lines for the deformable ones with $\Ca=0.2$, with the solid ---, dashed $--$ and dash-dotted $-\cdot-$ lines representing the fluid viscous stress, the particle viscous stress and the particle hyper-elastic contribution.}
  \label{fig:tauij}
\end{figure}
\begin{figure}[]
  \centering
  \input{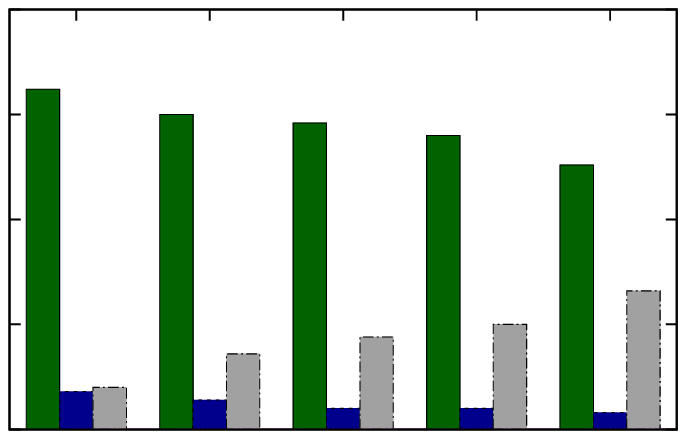}\\
  \vspace{0.55cm}
  \input{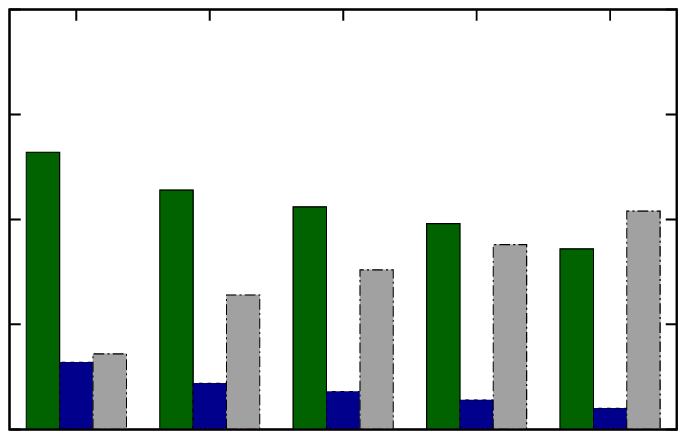}\\
  \vspace{0.55cm}  
  \input{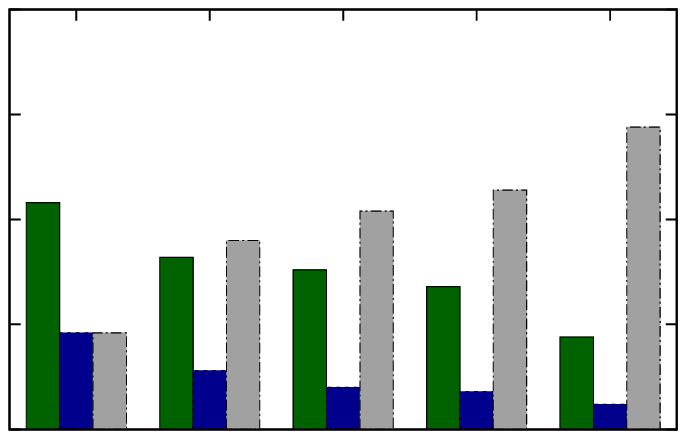}\\
  \vspace{0.55cm}
  \caption{Percentage contribution to the total shear stress $\bra{\bra{\sigma_{12}}}$ as a function of the Capillary number. The three panels correspond to different volume fractions: $\Phi=0.11$ (top panel), $0.22$ (middle panel) and $0.33$ (bottom panel). In each plot, the green, blue and grey bars represent the fluid viscous stress, the particle viscous stress and the particle hyper-elastic contribution, respectively.}
  \label{fig:balance}
\end{figure}

The total shear stress, $\sigma_{12}$, can be decomposed into the sum of the fluid $\sigmaf_{12}$ and particle stress $\sigmas_{12}$, see \equref{eq:phi-stress}. The mean fluid stress is the sum of the viscous and Reynolds stresses, while the solid one is the some of the viscous, Reynolds and hyper-elastic stresses:
\begin{multline}
\bra{\sigma_{12}} = \underbrace{\bra{ \left( 1 - \phi \right) \left( \muf \frac{du}{dy} - \rho u'v' \right)}}_{\rm fluid~stress} + \\
+ \underbrace{ \bra{\phi \left( \mus \frac{du}{dy} - \rho u'v' + G B_{12} \right)}}_{\rm solid~stress}.
\end{multline}
Each of these contribution have been averaged in time and in the homogeneous directions and displayed in \figrefS{fig:tauij} as function of the wall-normal distance $y$ for the cases with $\Ca=0.2$ and volume fractions $\Phi=0.11$, $0.22$ and $0.33$ (top, middle and bottom panel). The stresses are normalized by the total wall value. In the figure, the solid, dashed and dash-dotted lines represent the fluid viscous stress, the particle viscous and hyper-elastic stresses, respectively. The Reynolds stress contributions are negligible in all the cases considered here, thus not shown in the graphs. The circles and triangles are the fluid viscous stress and particle stress in the rigid case ($\Ca=0$) obtained from a complementary simulation.

We observe that the fluid viscous stress is the only not null component at the wall. At the lowest volume fraction shown here the fluid viscous stress is the dominant contribution, being responsible for more than $50\%$ of the total stress at each wall-normal location. It has a minimum value around $y\approx0.75h$ corresponding to the location of maximum particle stress and to the maximum particle concentration as previously discussed (see \figrefS{fig:velVof}). As the volume fraction is increased, the fluid viscous stress becomes smaller and smaller; this is compensated by an increase in the particle stress contribution which eventually becomes the dominant one. This behavior is observed for both rigid and deformable particles, with the main difference being a lower particle stress contribution across the all channel in the deformable case than the rigid one, especially close to the center line. Moreover, in the deformable particle suspension, we can further separate the particle stress in its viscous and hyper-elastic contributions. The hyper-elastic contribution is the dominant one, being responsible for almost the totality of the total stress and for its distribution across the domain. On the other hand, the viscous stress in the solid is almost null at low volume fractions, and progressively grows with it. Also, its profile is almost uniform across the whole domain, except close to the wall where it is always negligible as the solid volume fraction vanishes.

In \figrefS{fig:balance} we summarize the stress balance by showing the volume-averaged percentage contribution of all the non-zero components of the total shear stress, \ie fluid viscous stress (green), particle viscous stress (blue), and particle hyper-elastic stress (grey). Each panel shows how the stress balance change with Capillary number, and each panel corresponds to a different volume fraction ($\Phi=0.11$: top panel; $0.22$: middle panel; $0.33$: bottom panel). For every volume fraction, both the fluid and particle viscous stresses decrease with the Capillary number, while the particle elastic contribution increases (in the range of $\Ca$ considered here). Interestingly, notwithstanding the fact that the solid viscosity and volume fraction are constant, the normalized particle elastic stress decreases as $\Ca \rightarrow 0$, and the viscous contribution becomes relevant; conversely, as $\Ca \rightarrow \infty$ the percentage contribution of the particle viscous stress vanishes and the total particle stress reduces to its elastic contribution. Comparing the budget for varying volume fraction $\Phi$, we observe that as the volume fraction increases, the fluid stress progressively decreases due to the reduced volume of fluid, while the particle stress increases due to the increase of the solid fraction. The latter increase in the particle stress holds for both the viscous and elastic part of the particle stress, and is related to the increase of the number of particles in the domain.

\subsection*{Universal scaling of the effective viscosity}
\begin{figure}[h]
  \centering
  \input{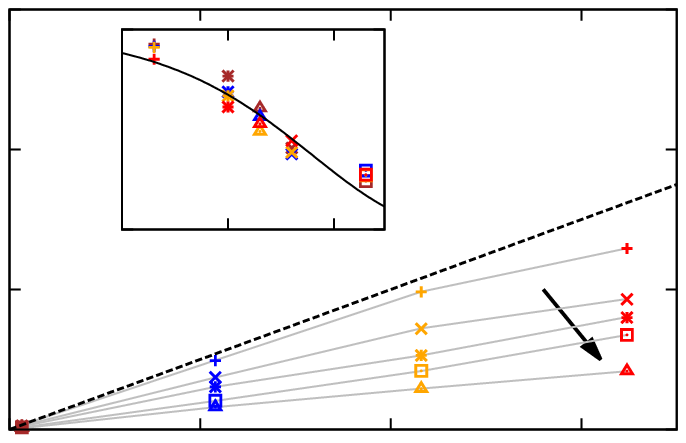}\\
  \vspace{0.55cm}
  \input{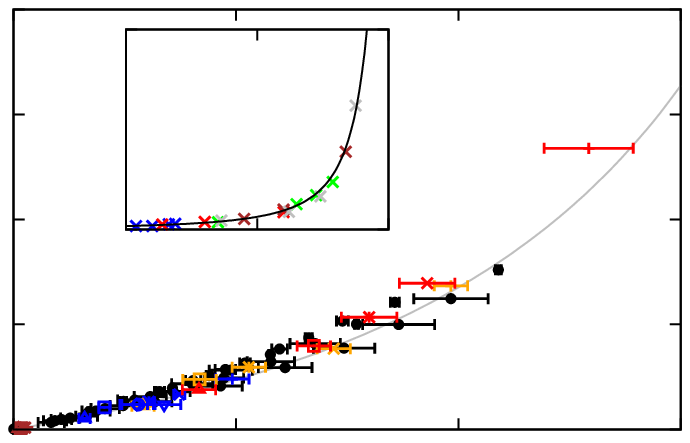}
  \vspace{0.55cm}
  \caption{(top) Effective volume fraction $\Phie$ as a function of the volume fraction $\Phi$. The black dashed line in the figure shows the profile $\Phie=\Phi$. The inset shows the ratio $\Phie/\Phi$ as a function of $\Ca$. (bottom) Effective viscosity $\mu$ divided with $\muf$ of all the considered cases as a function of the reduced effective volume fraction $\Phie$. The color and symbol scheme is the same as in the previous figures, with the addition of the black symbols which are the data of Ref.~\cite{matsunaga_imai_yamaguchi_ishikawa_2016a}. The error bars are computed based on the standard deviation of the Taylor parameter in the domain. The inset shows a similar fit, obtained with the red blood cell data from Ref.~\cite{dintenfass_1968a}.}
  \label{fig:scaling}
\end{figure}

Next, we propose a scaling for the suspension viscosity. As already discussed, the effective viscosity of rigid particle suspensions with negligible inertial effect is well described by the Eilers fit (\equref{eq:Eilers}), valid for a wide range of volume fractions. In the deformable case we can evaluate an effective volume fraction $\Phie$, a concept successfully used in the past for suspensions with different properties, such as charged colloidal particles, fiber and platelets suspensions, polyelectrolyte solutions \cite{mewis_frith_strivens_russel_1989a, frith_dhaene_buscall_mewis_1996a, quemada_1998a, mewis_wagner_2012a}. Here, we evaluate it based on spheres of radius equal to the  semi-minor axis of the ellipsoid $a$, the same used to compute the Taylor parameter, \ie $\Phie={\rm N} \times 4/3 \pi a^3/\mathcal{V}^{\rm tot}$ where ${\rm N}$ is the number of particles in the computational box of volume $\mathcal{V}^{\rm tot}$. Obviously, the effective volume fraction $\Phie$ is less than the nominal one $\Phi$, and thus represents a reduced volume fraction due to the particle deformations. We propose to define the effective volume fraction using the minor axis, differently from what done in previous works for fiber or rigid non-spherical suspensions \cite{batchelor_1971a, kerekes_2006a, lundell_soderberg_alfredsson_2011a}; the choice is motivated by the fact that the deformable particles are not rotating, thus, the excluded volume effect is less than the one based on the major axis; moreover, since the particles are approximately aligned with the mean shear direction, what affects the effective viscosity is the dimension in the direction normal to the mean shear, \ie the minor axis. As shown in the top panel of \figrefS{fig:scaling}, $\Phie$ increases with the volume fraction $\Phi$ and decreases with the Capillary number $\Ca$, similarly to the effective viscosity $\mu$; in particular, $\Phie$ is an almost linear and monotonic function of $\Phi$ and their ratio, independent of $\Phi$, decreases with $\Ca$ (see the inset figure). A fit to the data yields 
\begin{equation} \label{eq:phie}
\Phie = \Phi e^{-1.25\sqrt{\Ca}}.
\end{equation}
Note that the form of \equref{eq:phie} is chosen to satisfy the following conditions: \textit{i)} $\Phie$ is  a linear function of $\Phi$ for a fixed $\Ca$; this implies that deformability is weakly influenced by particle-pair interactions, typically proportional to $\Phi^2$. \textit{ii)} for $\Ca \rightarrow 0$ we expect to recover the rigid case with the effect of the deformability disappearing, \ie $\Phie = \Phi$. \textit{iii)} for $\Ca \rightarrow \infty$ the particles deform more and more, eventually giving no resistance to the fluid, \ie $\Phie = 0$. This simple relation is reported in the inset of the top panel in \figrefS{fig:scaling} to show the quality of the fitting. Note that, \equref{eq:phie} has been derived for the case with $K=1$, thus no explicit dependency on the viscosity ratio has been included.

Next, we plot the effective viscosity $\mu$, already shown in \figrefS{fig:viscFG}, now as a function of $\Phie$, see the bottom panel of \figrefS{fig:scaling}. As clearly seen, all the data collapse on an universal curve, which is well described by the Eilers fit used for the rigid-particle suspensions. Note that, this scaling applies to all the volume fractions $\Phi$ and Capillary numbers $\Ca$ considered (the fit works also for all the cases with different solid to fluid viscosity ratio $\Key$ considered in the present work which will be discussed later). Moreover, we also include in the figure the data taken from Ref.~\cite{matsunaga_imai_yamaguchi_ishikawa_2016a} pertaining a suspension of capsules (black symbols) as a proof of the universality of the scaling. This is further proved in the inset figure, where the red blood cell measurement of Ref.~\cite{dintenfass_1968a} for different $\Ca$ and $\Key$ are collapsed on the Eilers formula; note that, in order to apply our result to these data, we first fit the Eilers formula to the case of the rigid RBCs, obtaining $B = 1.25$ and a maximum packing fraction $\Phi_m = 0.88$, as the undeformed shape of the RBCs is not spherical but disk-like \cite{mueller_llewellin_mader_2009a}. 

Therefore, we have shown that the effect of the particles deformation can be included into the suspension shear stress as follow
\begin{equation}
\sigma_{12}=\muf \dot{\gamma} \mathcal{F} \left( \Phie \left( \Phi, \Ca, \Key \right) \right),
\end{equation}
where $\Phie$ is the measure of the reduction in the nominal volume fraction due to the deformation, (which we have estimated with \equref{eq:phie} for the case of $\Key=1$), and $\mathcal{F}$ is the Eilers formula in \equref{eq:Eilers} or any other approximation of the effective suspension viscosity as function of the solid volume fraction. In summary, this closure provides the shear stress for non-Newtonian suspensions made of deformable elastic particles, with microstructure effects coded into the analytical function $\mathcal{F}$. This closure is valid for suspension flows with negligible inertia.

\begin{figure}[t]
  \centering
  \input{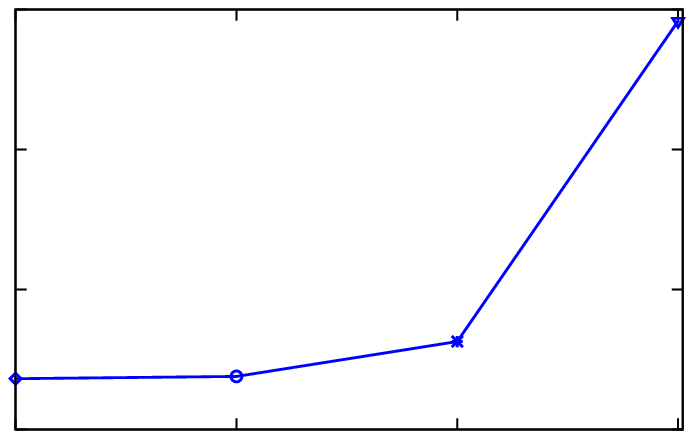}\\
  \vspace{0.55cm}
  \input{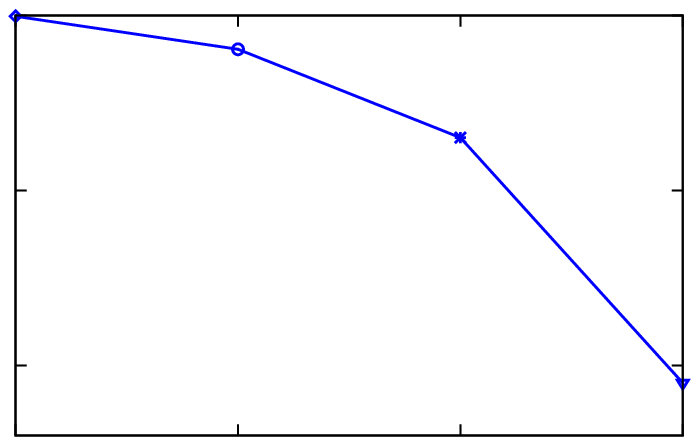}\\
  \vspace{0.55cm}
  \caption{(top) Normalised effective viscosity $\mu$, (bottom) Taylor parameter $\Tay$ as a function of the solid to fluid viscosity ratio $\Key$. The data refer to the volume fraction $\Phi=0.11$ and Capillary number $\Ca=0.2$.}
  \label{fig:K}
\end{figure}

\subsection*{Effect of the solid to fluid viscosity ratio $\Key$}
\begin{figure}[t]
  \centering
  \input{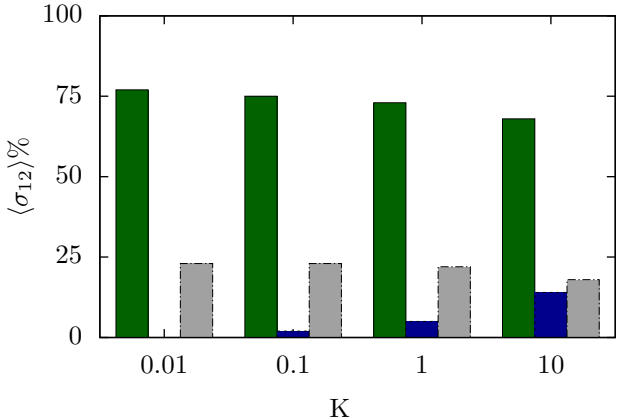}\\
  \vspace{0.55cm}  
  \caption{Shear budget balance as a function of the solid to fluid viscosity ratio $\Key$. The data refer to the volume fraction $\Phi=0.11$ and Capillary number $\Ca=0.2$.}
  \label{fig:Khist}
\end{figure}
Finally, in this last section, we briefly assess the effect of the parameter $\Key$, \ie the ratio of the solid viscosity $\mus$ and the fluid one $\muf$. Here, we focus our analysis on a single volume fraction $\Phi=0.11$ and Capillary number $\Ca=0.2$, and we consider four different values for $\Key$: $0.01$, $0.1$, $1$ (also discussed previously) and $10$.

The two panels in \figrefS{fig:K} shows the effect of $\Key$ on the effective viscosity $\mu/\muf$ (top) and on the Taylor parameter $\Tay$ (bottom). The effective viscosity monotonically increases with $K$, changing approximately from $1.24$ to $1.5$ due to the increase of $\Key$ by a factor of $1000$. This suggest a weak dependence of the suspension viscosity on this parameter (compared to the one with $\Ca$), at least for the considered volume fraction. Differently, the Taylor parameter $\Tay$ decreases for an increase of $\Key$. These results suggest that high values of $\Key$ make the particle effectively more rigid, thus reducing the overall deformation ($\Tay$), resulting in higher suspension viscosities; on the other hand,  low values of $\Key$ increase the particles deformation ($\Tay$), thus reducing the suspension viscosity. The limit behavior for $\Key \rightarrow \infty$ is the rigid particle case, while for $\Key \rightarrow 0$ the particle is more deformable, but the particle contribution to the suspension viscosity does not vanish completely due to the finite Capillary number $\Ca$, \ie $\mu/\muf \nrightarrow 1$ for $\Key\rightarrow0$; this is different from what observed for the Capillary number $\Ca$ where $\mu/\muf \rightarrow 1$ for $\Ca\rightarrow \infty$. Finally, \figrefS{fig:Khist} shows the total shear budget, similarly to \figrefS{fig:balance}. We note that the fluid stress slightly decreases with $\Key$, thus the total particle stress slightly increases. This is obtained by a strong increase of the viscous particle stress, and contrasted by a small reduction of the elastic counterpart. Similarly to the discussion above, we observe that low values of $\Key$ only slightly modify the suspension rheology, while changes can be noticed for high values of $\Key$ when the particle become more rigid and less deformable, thus approaching the behaviour of rigid spheres.

\section{Conclusion} \label{sec:conclusion}
We have studied the rheology of a suspension of deformable viscous hyper-elastic particles in a Newtonian fluid in a wall-bounded shear flow, \ie Couette flow, at low Reynolds number such that inertial effects are negligible. The deformable particles are made of a neo-Hookean material, satisfying the Mooney-Rivlin law. The multiphase flow is solved with the use of a one-continuum formulation by introducing an indicator function to distinguish the fluid and solid phases, \ie the solid volume fraction $\phis$. The results are numerically obtained by solving the conservation of momentum and the incompressibility constraint in a fully Eulerian fashion.

The rheology of the suspension is analyzed by discussing how the suspension effective viscosity $\mu$ is affected by variations of the particle volume fraction $\Phi$, the Capillary number $\Ca$ and the solid to fluid viscosity ratio $\Key$. We observed that $\mu$ is a non-linear function of all this parameters $\mu=\mu \left( \Phi, \Ca, \Key \right)$, being $\Phi$ and $\Ca$ the most effective. The suspension of deformable particles has a viscosity lower than the one for rigid particles; this is due to the deformation, quantified here by the Taylor parameter $\Tay$, which grows with both $\Phi$ and $\Ca$, but decreases with $\Key$. As the Capillary number is proportional to the shear, suspensions of deformable particles are shear-thinning, as known for blood. The non-linear dependency on the different parameters is further exemplified by the first and second normal stress differences, $\NSI$ and $\NSII$. These are not null, indicating the visco-elasticity nature of the suspension, and show a non-monotonic dependence with $\Ca$. $\NSI$ is positive while $\NSII$ negative, with $\vert \NSI \vert > \vert \NSII \vert$, similarly to what found for suspension of flexible filaments and polymers. With a stress budget study, we have shown that the particle stress grows with both the volume fraction $\Phi$ and the Capillary number $\Ca$, and that the particle stress is determined by its elastic part, with only a weak viscous contribution, which exhibits an opposite trend with $\Ca$.

Finally, we propose an universal scaling for the effective viscosity of suspensions of deformable particles, which is able to collapse all the data onto the Eilers fit, usually valid for rigid particles. To this end, we introduce a reduced effective volume fraction, function of the capillary number and of the nominal volume fraction, accounting for the effect of deformability. Based on our and others' data, we provide an estimate of this effective volume fraction and hence an analytical closure for the shear stress valid for suspension of deformable particles (and capsules) with negligible inertia.

This work proposes a new approach to suspensions of deformable objects and can be extended in a number of ways. Additional simulations and experiments may improve the estimate given here of the effective volume fraction by quantifying particle deformation for \eg particles of different shapes, such as oblate or biconcave red-blood cells. Alternatively, one may consider inertial effects as in Ref.~\cite{picano_breugem_mitra_brandt_2013a}, and the effect of a time-dependent shear rate to relate the particle average deformation  to the history of the applied stresses. For slow variations of shear rates, however, we expect to retrieve the rheological properties described here, \ie shear thinning, while memory effect may appear for fast enough deformations.

\section*{Acknowledgment}
The work of MER and LB was supported by the European Research Council grant no.\ ERC-2013-CoG-616186, TRITOS and by the Swedish Research Council (grant no.\ VR 2014-5001). DM is supported by grants from the Swedish Research Council (grant no.\ 638-2013-9243 and 2016-05225). The authors acknowledge computer time provided by SNIC (Swedish National Infrastructure for Computing), and fruitful discussions with Prof.\ Dhrubaditya Mitra (NORDITA), Dr.\ Sarah Hormozi (Ohio University) and Dhiya Abdulhussain Jassim Alghalibi (KTH) for the data pertaining rigid particle suspensions.

\section*{References}
\bibliographystyle{elsarticle-num}
\bibliography{./bibliography.bib}
\end{document}